\documentclass[12pt]{iopart}

\usepackage[utf8]{inputenc}  %Input what you want e.g., é, ł, a, ü
\usepackage[T1]{fontenc}     %Output what you want e.g., é, ł, a, ü
\usepackage[british]{babel}  %Do hyphenation according to british english
\usepackage[colorlinks,linkcolor=magenta,urlcolor=blue]{hyperref}  %Hyperlinks (pink, green, blue)
\usepackage{graphicx} % Package to insert exteral figures
\usepackage{iopams}
\usepackage{dsfont}
\usepackage{xspace}  %Useful to add space in macros
\usepackage{centernot}
\usepackage{pgfplots}  %Useful to plot graphs with latex
\newcommand{\bra}[1]{\left\langle #1 \right|}
\newcommand{\ket}[1]{\left| #1 \right\rangle}

\newcommand{\ketbra}[2]{\left|#1\middle\rangle\middle\langle#2\right|}

\def\be{\begin{eqnarray}}
\def\ee{\end{eqnarray}}
\newcommand{\eqref}[1]{(\ref{#1})}
\newcommand{\figref}[1]{Figure~\ref{#1}}

\begin{document}

\title{Autonomous quantum thermal machine for generating steady-state entanglement}

\author{Jonatan Bohr Brask$^{1}$\footnote{These authors contributed equally to this work.}, G\'eraldine Haack$^{2,3}$\footnotemark[1], Nicolas Brunner$^1$, Marcus Huber$^{4,5}$}

\address{$^1$ Département de Physique Théorique, Université de Genève, 1211 Genève, Switzerland}
\address{$^2$ Univ. Grenoble Alpes, INAC-SPSMS, F-38000 Grenoble, France}
\address{$^3$ CEA, INAC-SPSMS, F-38000 Grenoble, France}
\address{$^4$ Departament de F\'isica, Universitat Aut\`onoma de Barcelona, E-08193 Bellaterra, Spain}
\address{$^5$ ICFO-Institut de Ciencies Fotoniques, Mediterranean Technology Park, 08860 Castelldefels (Barcelona), Spain}
\ead{jonatan.brask@unige.ch}

\date{\today}  

\begin{abstract}
We discuss a simple quantum thermal machine for the generation of steady-state entanglement between two interacting qubits. The machine is autonomous in the sense that it uses only incoherent interactions with thermal baths, but no source of coherence or external control. By weakly coupling the qubits to thermal baths at different temperatures, inducing a heat current through the system, steady-state entanglement is generated far from thermal equilibrium. Finally, we discuss two possible implementations, using superconducting flux qubits or a semiconductor double quantum dot. Experimental prospects for steady-state entanglement are promising in both systems. 
\end{abstract}

\maketitle

\section{Introduction}

The generation of entangled states in quantum systems represents a central challenge for quantum information processing and fundamental tests of quantum theory. Tremendous progress has been achieved in particular with the development of methods to efficiently counter various (and essentially unavoidable) sources of noise, such as coupling to the environment. Recently, it was realized that noise and coupling to the environment are not always detrimental, and can be used advantageously in certain situations \cite{plenio99,kim,lech,braun,benatti,burgarth,bellomo}. While these schemes allow only for transient entanglement,  it was shown that steady-state entanglement can be obtained from dissipative processes \cite{diehl,verstraete,kraus,ticozzi2}. The creation of steady-state entanglement was investigated for trapped ions \cite{milburn}, atoms in cavities \cite{kastoryano,wang}, superconducting \cite{reiter13} and spin qubits \cite{schuetz,Cai2010}, and nanomechanical systems \cite{jens}, with experimental implementations reported \cite{krauter,barreiro,lin,shankar}. The main ingredients are engineered decay processes and quantum bath engineering \cite{beige,kastoryano2,aron}. These approaches drive the system into a single fixed point corresponding to an entangled state, but require an external coherent driving field, which can be considered a source of work. 

It is thus natural to ask if steady-state entanglement can be generated via incoherent interactions with thermal environments alone. Indeed, this can be achieved in a situation of thermal equilibrium, by placing a system featuring entanglement in the ground state in thermal contact with a cold bath. More interestingly it was shown that steady-state entanglement can be generated far from equilibrium, without using any source of coherence or external control. The first example considered an atom coupled to two cavities and driven by incoherent light \cite{plenio}. Subsequent work discussed this problem in the context of many-body systems \cite{hartmann1,hartmann2}, interacting spins \cite{znidaric,quiroga07}, and atoms placed in a thermal environment \cite{bellomo1,bellomo2}. Steady-state entanglement was also shown to be beneficial for transport \cite{manzano} and cooling \cite{brunner14}. More generally, the potential of thermal entanglement generation is still not well understood. In particular, coupling to a thermal environment is arguably the most common and natural source of dissipation, hence using it advantageously may lead to novel experimental possibilities in the context of quantum information, and might also give insight into possible generation of entanglement in biological systems \cite{nori}.

Here we discuss this problem in what is arguably the simplest possible setting: two resonant qubits, each in weak thermal contact with a heat bath. Placing the two heat baths at different temperatures results in a net heat current passing through the system, which can generate steady-state entanglement far from thermal equilibrium. Our setup makes use of a source of free energy (i.e. two heat baths at different temperatures) and can thus be considered a thermal machine for generating steady-state entanglement. The machine is autonomous in the sense that it uses only incoherent interactions with thermal baths, but no source of coherence or external control. We investigate the amount of entanglement that can be generated with respect to the coupling parameters and temperatures of the baths. Then we illustrate the practical relevance of our model by discussing two implementations in superconducting flux qubits \cite{Mooij99}, and in a semiconductor double quantum dot \cite{Wiel03}. Thanks to their coherence properties and high tunability in the quantum regime, these systems are natural candidates to test the limits of dissipation processes as a resource for steady-state entanglement. \\

\begin{figure}[t]
\centering
\includegraphics[width=0.6\linewidth]{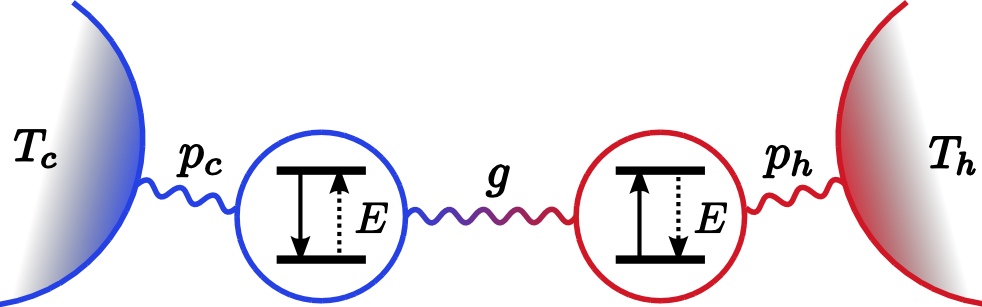}
\caption {Sketch of the quantum thermal machine.}
\label{fig.twoqubitscheme}
\end{figure}

\section{Model}

We consider two qubits with identical energy gaps $E$ weakly coupled to each other and to separate thermal reservoirs (\figref{fig.twoqubitscheme}). We denote the ground and excited states $\ket{0}$, $\ket{1}$, and the free Hamiltonian for the qubits in this basis is
\begin{equation}
\label{eq.absHo}
\hat{H}_0 = E ( \ketbra{1}{1}\otimes\mathds{1} + \mathds{1}\otimes\ketbra{1}{1} ) ,
\end{equation}
The interaction Hamiltonian, which is energy conserving, is given by
\begin{equation}
\label{eq.absHint}
\hat{H}_{int} = g ( \ketbra{10}{01} + \ketbra{01}{10} ),
\end{equation}
where $g$ is the strength of the coupling between the qubits. The coupling to the thermal baths is modelled using a simple collision model  where thermalisation happens through rare but strong events. At every time step, each qubit $k$ is either reset to a thermal state $\tau_k$ at the temperature of its bath with a small probability or left unchanged. The state of the qubits evolves according to the master equation
\begin{equation}
\label{eq.twoqubitmaster}
\frac{\partial\rho}{\partial t} = i [\rho,\hat{H}_0 + \hat{H}_{int}] + \sum_{k \in \{c,h\}} p_k (\Phi_k(\rho) - \rho)
\end{equation}
where $p_k$ is the thermalisation rate for qubit $k$ and $\Phi_c(\rho) = \tau_c\otimes\Tr_c(\rho)$ and $\Phi_h(\rho) = \Tr_h(\rho)\otimes\tau_h $. We take the first qubit to have the colder and the second to have the warmer bath temperature. We refer to them as the 'cold' and 'hot' qubit respectively and use subscripts $c$ and $h$. The thermal states are given by $\tau_k = r_k \ketbra{0}{0} + (1-r_k)\ketbra{1}{1}$ with occupation probabilities determined by the Boltzmann factor according to $r_k = 1/(1+e^{-E/T_k})$ where $T_k$ is the reservoir temperature for qubit $k$ (throughout the paper we set $k_B = 1$ and $\hbar=1$). Note that the master equation applies in the perturbative regime $p_c,p_h,g \ll E$ and $p_c,p_h \ll 1$ \cite{skrzypczyk2011}.

Next we look for the steady-state solution of \eqref{eq.twoqubitmaster}. Since \eqref{eq.twoqubitmaster} is linear in $\rho$, it can be recast as a matrix differential equation $\frac{\partial \mathbf{v}}{\partial t} = A \mathbf{v} + \mathbf{u}$, where $\mathbf{v}$ is a rewrapping of the density matrix $\rho$ to a vector, and the matrix $A$ and vector $\mathbf{u}$ depend on $E$, $g$, $p_k$, $T_k$, and encode the right-hand side of \eqref{eq.twoqubitmaster}. The steady-state solution is given by $\mathbf{v}_\infty = - A^{-1}\mathbf{u}$. Wrapping back to matrix form, we obtain the steady-state density matrix
\begin{equation}
\label{eq:rho_infty}
\rho_\infty = \gamma \bigg[ p_cp_h \tau_c\otimes\tau_h + \frac{2g^2}{(p_c+p_h)^2} (p_c\tau_c + p_h\tau_h)^{\otimes 2} + \frac{gp_cp_h(r_c-r_h)}{p_c+p_h} \mathcal{Y} \bigg]
\end{equation}
with $\mathcal{Y} = i\ketbra{01}{10} - i\ketbra{10}{01}$ and $\gamma = 1/(2g^2 + p_cp_h)$, and where $\rho^{\otimes 2}=\rho \otimes \rho$. Note that for resonant qubits, the steady state depends on the energy $E$ only through $r_c$, $r_h$. 
We also determine the heat currents in the system. The energy flowing from qubit $k$ to its bath is given by the product of the thermalisation rate and the change in energy of the qubit at each thermalisation event
\begin{equation}
Q_k(\rho) = p_k E \bra{1} (\rho_k - \tau_k) \ket{1} ,
\end{equation}
where $\rho_k$ is the reduced state for qubit $k$ corresponding to the joint state $\rho$.

\begin{figure}[t]
\centering
\includegraphics[width=0.995\linewidth]{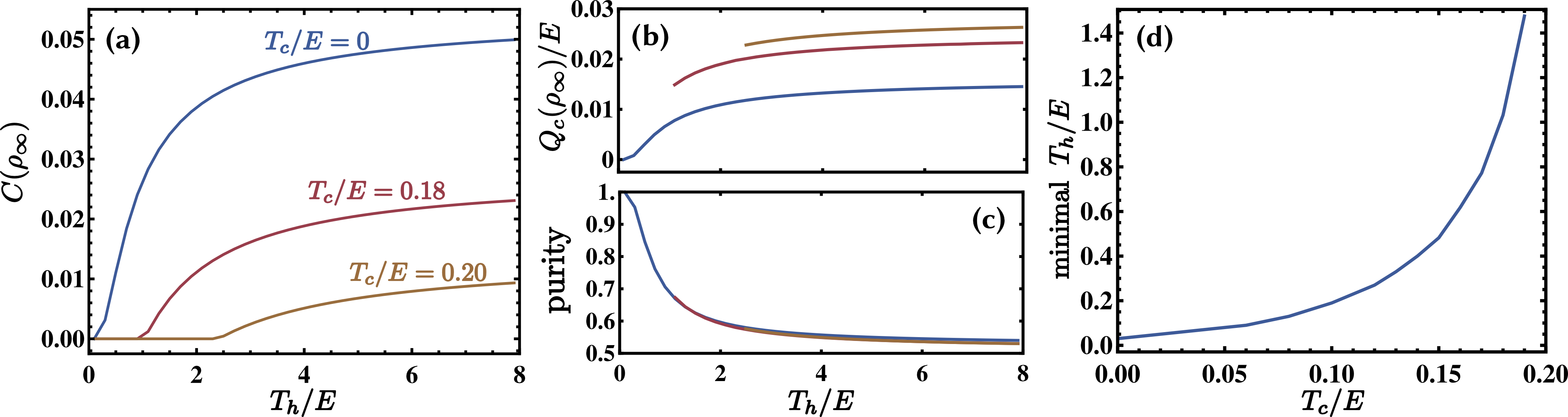}
\caption {(Color online) Characterization of steady-state entanglement \textbf{(a)} Concurrence vs.~hot bath temperature $T_h$, for different cold bath temperatures $T_c$. \textbf{(b)} Heat current $Q_c(\rho_\infty)$ from the cold qubit to its bath, and \textbf{(c)} purity corresponding to the parameter settings in (a). \textbf{(d)} Threshold hot bath temperature required to generate entanglement in the steady state as a function of the cold bath temperature $T_c$.}
\label{fig.plots_absmodel}
\end{figure}

We are now in position to discuss the entanglement of the steady state. As a measure of entanglement, we use the concurrence \cite{wootters}, which for the steady state (\ref{eq:rho_infty}) can be written
\begin{equation}
\label{eq.twoqubitconcsteady}
C(\rho_\infty) = \textrm{max} \left\{ 0, f(r_c,r_h) - \sqrt{h(r_c,r_h)h(1-r_c,1-r_h)} \right\}
\end{equation}
with
\begin{eqnarray}
f(r_c,r_h) & = \gamma \frac{gp_cp_h}{p_c+p_h}|r_c-r_h| , \\
h(r_c,r_h) & = \gamma \left( p_cp_hr_cr_h + 2g^2 \left(\frac{p_cr_c + p_hr_h}{p_c+p_h}\right)^2 \right) .
\end{eqnarray}
Notice that when the two temperatures coincide, i.e. $T_c=T_h$, we have $C(\rho_\infty)=0$ since $f(r_c,r_h)=0$ in this case. That is, at equilibrium the steady state of the two qubits is always separable. However, when moving away from equilibrium by choosing different temperatures for the two baths, hence establishing a heat current from the hot to the cold bath, steady-state entanglement can be generated as we will now see.

We first discuss the case $T_c = 0$. For any $T_h>0$, a heat current is created and steady state entanglement appears. The top curve in \figref{fig.plots_absmodel}(a) shows the maximal amount of entanglement that can be achieved as a function of $T_h$ by optimising the coupling parameters (with the constraint that $g,p_c,p_h < 10^{-2}$ to ensure the validity of our master equation). The corresponding heat current $Q_c(\rho_\infty)$ is plotted in \figref{fig.plots_absmodel}(b). It is clearly seen that increasing $T_h$, hence increasing the heat current, creates more entanglement. The largest amount of entanglement, $C(\rho_\infty) \approx 0.054$, is obtained when $T_h \rightarrow \infty$ and $g \approx 1.6\times 10^{-3}$, $p_c \approx 10^{-2}$, $p_h \approx 1.1\times 10^{-3}$.
Next we consider the case $T_c>0$. In this case a minimal temperature difference (and thus a minimal heat current) is required to get entanglement, as is apparent from \figref{fig.plots_absmodel}(a). The threshold hot bath temperature depends on $T_c$ (see \figref{fig.plots_absmodel}(d)), and above $T_c/E \approx 0.21$ no entanglement can be generated. We also computed the purity of the steady state, given by $\tr(\rho_\infty^2)$, which depends on $T_h$ but is essentially independent of $T_c$ (see \figref{fig.plots_absmodel}(c)). 

The simplicity of the above model makes it rather versatile, we believe. Notably, we considered fully incoherent coupling to the heat baths, and made no assumption about the structure of these baths. This will be illustrated in the next sections where we discuss two possible implementations.\\

\begin{figure}[t]
\centering
\includegraphics[width=0.6\linewidth]{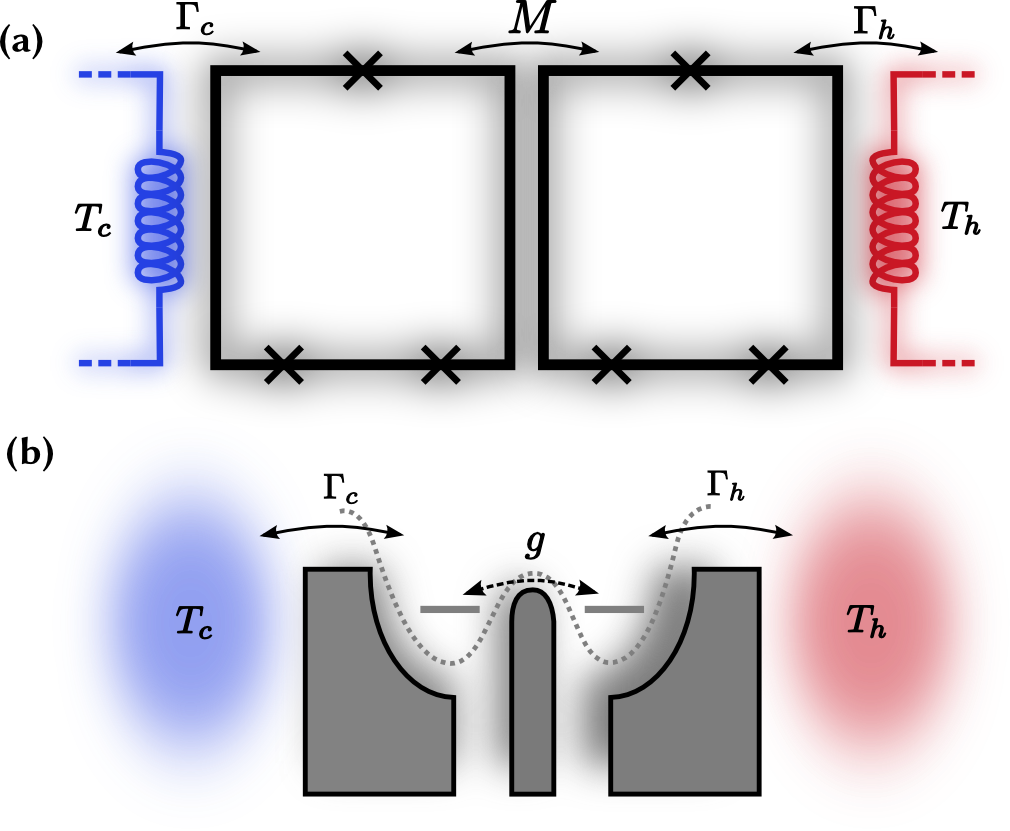}
\caption {Implementations of our model in \textbf{(a)} superconducting flux qubits, and \textbf{(b)} a double quantum dot.}
\label{fig.impl}
\end{figure}

\section{Implementation 1}

The high-tunability of superconducting flux qubits has motivated lots of experiments, demonstrating for instance the control of the inductive coupling between two flux qubits \cite{Majer05}, the preparation of entangled states \cite{Izmalkov04}, and the observation of the ultrastrong coupling regime between a flux qubit and its environment \cite{Forn-Diaz10}. Here, we consider a circuit made of two flux qubits which interact via a shared inductance $M$ as shown in \figref{fig.impl}(a)). When appropriately tuned \cite{Liu,Xia}, this system is described by the Hamiltonian $\hat{H}_{flux} = \hat{H}_0 + \hat{H}_{int}$, c.f. Eqs. \eqref{eq.absHo} and \eqref{eq.absHint}, where $g$ is the interaction strength now set by $M$ (see \ref{app.impl1}). Moreover, each qubit is subject to dissipation processes which simply arises from the finite impedance of external coils required to operate the two superconducting loops as effective two-level systems. These external circuits are characterized by noise spectra $S_k(\omega)$ which depend on the Bose-Einstein distribution $n_B(\omega, T) = 1/(e^{ \omega/ T} -1)$. Tuning the noise of the external circuit enables therefore the control of the temperature of the environments coupled to each qubit independently. We describe the interaction between each flux qubit and its dissipative environment via an Hamiltonian of the form \cite{Martinis03}
\be
\label{eq.Hbath}
\hat{H}_\textrm{q-e}= \sum_{k \in \{c,h\}} \sqrt{\Gamma_k} \,  \hat{i}_k \, \Big(\vert 1 \rangle_k \langle 0 \vert + \vert 0 \rangle_k \langle 1 \vert \Big)\,,
\ee
where $\hat{i}_k$ is the fluctuating current in external circuit $k$. Standard quantum optics calculations \cite{Breuer, Schaller} allows us to derive the Lindblad equation governing the dynamics of the interacting flux qubits in presence of their thermal environments
\begin{eqnarray}
\label{eq.master}
\frac{\partial \rho}{\partial t} =  i [\rho,&\hat{H}_{flux} ] + \sum_{k \in \{c,h\}} \Gamma_k^+  \left( \hat{J}_k \rho \hat{J}_k^\dagger - \frac{1}{2} \left\{ \hat{J}_k^\dagger \hat{J}_k, \rho \right\} \right) \nonumber \\
& +  \sum_{k \in \{c,h\}} \Gamma_k^-  \left( \hat{J}_k^\dagger \rho \hat{J}_k - \frac{1}{2} \left\{ \hat{J}_k \hat{J}_k^\dagger, \rho \right\} \right)\,.
\end{eqnarray}

We note that this equation is similar to \eqref{eq.twoqubitmaster}. The jump operators $\hat{J}^\dagger_k$ and $\hat{J}_k$  correspond to the raising and lowering operators for each qubit, $\hat{J}_c = \vert 1 \rangle \langle 0 \vert \otimes \mathds{1}$ and $\hat{J}_h = \mathds{1} \otimes  \vert 1 \rangle \langle 0 \vert$. The process corresponding to qubit $k$ absorbing (emitting) an excitation is characterized by the rate $\Gamma_k^+$ ($\Gamma_k^-$), which is proportional to $n_B$ $ (1+ n_B)$. We refer the reader to the Appendix for more details on the derivation of Eq.~(\ref{eq.master}) and the form of the rates $\Gamma_k^{\pm}$.

\begin{figure}
\centering
\includegraphics[width=0.995\linewidth]{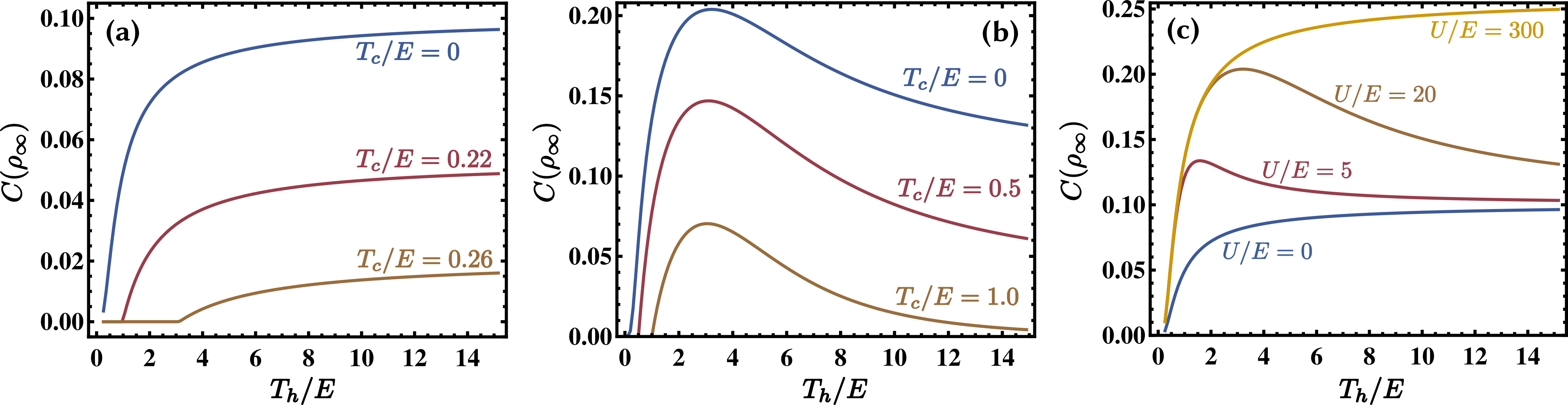}
\caption {(Color online) Steady-state entanglement for both proposed implementations. \textbf{(a)} Concurrence vs.~the temperature of the warmer bath $T_h$ for the flux-qubit system, for different cold bath temperatures $T_c$ as indicated. \textbf{(b)} Concurrence vs. $T_h$ for the double-quantum-dot system, for different $T_c$ as indicated and Coulomb energy $U/E=20$. \textbf{(c)} Concurrence vs. $T_h$ for the double-quantum-dot system, for the different Coulomb energies as indicated, and $T_c=0$ (note that the second curve from the top corresponds to the top curve in (b)).}
\label{fig.plots_impl}
\end{figure}

Applying the same techniques as for our simple model, we characterize the steady state of the system and study the entanglement between the flux qubits. \figref{fig.plots_impl}(a) shows the concurrence for different temperatures; again an optimization over the coupling parameters $g$, $\Gamma_c$, $\Gamma_h$ is performed (in the weak coupling regime) and the results are qualitatively very similar to those of our simple model, see \figref{fig.plots_absmodel}(a)). We find a maximal amount of entanglement of $C(\rho_\infty) \simeq 0.1$ and the threshold cold bath temperature below which steady-state entanglement is possible is $T_c/E \simeq 0.283$. Considering that transition frequencies of flux qubits are in the GHz range, this threshold cold bath temperature corresponds to few mK. Note that this temperature range actually corresponds to the typical temperatures at which those experiments are performed. This characteristic, as well as enhanced coherence times of the order of $10 \, \mu$s recently reported in Refs.~\cite{Bylander11, Stern14}, makes flux qubits promising candidates to realize the autonomous thermal machine we propose.\\

\section{Implementation 2}

Next we consider a double quantum dot tunnel-coupled to fermionic reservoirs characterised by temperatures $T_c$, $T_h$ and chemical potentials $\mu_c$, $\mu_h$ as shown in \figref{fig.impl}(b)). As shown in the Appendix, the Hamiltonian of this system takes the form $\hat{H}_{dot} + \hat{H}_{q-E}$ with $\hat{H}_{dot} = \hat{H}_0 + \hat{H}_{int}+ U \ket{11}\bra{11 }$. Compared to the flux qubits Hamiltonian, $\hat{H}_{dot}$ is characterized by an additional inter-dot interaction set by the Coulomb energy $U$. When the interaction between the dots and the reservoirs is weak, only single-charge tunnelling events occur and the dynamics of the system is captured by a master equation of the form \eqref{eq.master}. However, the rates $\tilde{\Gamma}_k^{\pm}$ now reflect the fermionic nature of the system, $ \tilde{\Gamma}_k^+  = \Gamma_k \,  n_F(E, T_k)$ and $\tilde{\Gamma}_k^- =  \Gamma_k \, \big( 1-n_F(E, T_k) \big)$ (see \ref{app.impl2}). Here $n_F(E,T) = 1/(e^{E/ T}+1)$ is the Fermi-Dirac distribution.

\figref{fig.plots_impl}(b) shows the concurrence in the steady state for varying temperatures and a fixed, non-zero, Coulomb energy. Again we observe a behaviour similar to the simple model (see \figref{fig.plots_absmodel}(a)). Interestingly, for $U \neq 0$, the temperature $T_h$ for which entanglement is maximised turns out to be finite. Indeed, the inter-dot Coulomb interaction prevents a second electron from one of the reservoirs from jumping into the system for a given range of $T_h$. This tends to increase the amount of entanglement generated.

Another key feature of this model is the dependance of the amount of entanglement that can be generated and of the threshold temperature on $U$, see \figref{fig.plots_impl}(c). For $U=0$ we find $C(\rho_\infty) \simeq 0.10$ (similarly to the flux qubit case) \footnote{Due to the fermionic nature of the systems, the value of the concurrence represents only a lower bound, as taking into account super-selection rules one expects to find more entanglement, see e.g.~\cite{caban}.}, while the largest amount of entanglement, $C(\rho_\infty) \simeq 0.2587$, is found for large $U \approx 25\times 10^3 E$. For $U=0$, the threshold cold bath temperature above which no entanglement can be found is the same as for the flux qubit system, $T_c/E \approx 0.283$, but the threshold can be made arbitrarily large by increasing $U$. E.g.~for $U=300$, the threshold is $T_c/E \approx 21.4$. Experimentally, semiconductor quantum dots are highly controllable thanks to external gate voltages that can be used to tune the different parameters in the desired range. This tunability, as well as coherence times on the order of ns \cite{Petersson10} with energy splitting $\sim$ 1 meV, makes experimental perspectives also promising for this system.\\

\section{Conclusion}

We discussed a model for an autonomous quantum thermal machine, able to generate steady-state entanglement between two interacting qubits. Remarkably, our scheme only relies on incoherent interactions with thermal baths. We proposed two implementations with widely investigated mesoscopic systems, one with two superconducting flux qubits and one with a double quantum dot. We considered relevant experimental values for the various parameters and obtained promising results for both platforms. Perspectives to this work concern the possibility to enhance the significant, but non-maximal, amount of entanglement generated by the model we propose. A first option could be to use entanglement distillation, a process which can be achieved dissipatively \cite{vollbrecht}. Another possibility is to look for schemes using higher dimensional quantum systems. The present model may also be relevant in the context of quantum biology, where the role of quantum coherence and entanglement is currently investigated. Going beyond the scope of generating steady-state entanglement, i.e.~entanglement available on demand, a promising direction concerns the transient regime. Just like in refrigeration schemes \cite{Mitch,inprep}, the finite time behaviour may lead to enhanced entanglement that can, together with precise timing, be extracted at regular intervals.

\ack

We thank Jukka Pekola for discussions on the physical implementations. We acknowledge financial support from the Swiss National Science Foundation (grant PP00P2\_138917 and Starting Grant DIAQ), SEFRI (COST action MP1006), and the EU SIQS. GH also acknowledges support from the ERC grant MesoQMC. MH acknowledges funding from the Juan de la Cierva fellowship (JCI 2012-14155), the European Commission (STREP "RAQUEL") and the Spanish MINECO Project No. FIS2013-40627-P, the Generalitat de Catalunya CIRIT Project No. 2014 SGR 966.

\appendix

\section{Implementation with flux qubits}
\label{app.impl1}

As in recent experiments, each flux qubit of our model consists of a superconducting loop with several Josephson junctions, see \figref{fig.impl}. The increased number of Josephson junctions makes the circuit less sensitive to magnetic flux noise \cite{You08}. The magnetic flux threading the loop induces clockwise and anti-clockwise supercurrents, $\pm I$, which define two classical states. When the magnetic flux is close to half a flux quantum, the eigenstates of the system are a superposition of the clockwise and anticlockwise super current states $\vert + I \rangle$ and $\vert -I \rangle$ and are well separated from higher energy levels. Hence, each circuit behaves as an effective two-level system characterized by an eigenenergy $\omega$ \cite{Wal00, You08, Stern14} (We set $\hbar = k_B = 1$ as in the main text). If the two qubits are close to each other, they interact via a shared inductance $M$. In the two-level basis and using the rotating wave approximation, the Hamiltonian of the two coupled flux qubits (labelled $c$ and $h$ as in the main text) reads \cite{Izmalkov04,Majer05,You05,Liu}
\be
\label{eq.Ham_flux}
\hat{H}_{flux} &=& \omega_c ( \vert 1 \rangle \langle 1 \vert \otimes\mathds{1}) + \omega_h (\mathds{1}\otimes \vert 1 \rangle \langle 1 \vert) \\
&&+ \lambda_1 (\vert 01 \rangle \langle 10 \vert + h.c. ) + \lambda_2 (\vert 00 \rangle \langle 11 \vert + h.c.) \,,\nonumber
\ee
with
\be
\lambda_1 &=& M \langle 10 \vert \hat{I}_c \otimes \hat{I}_h \vert 01 \rangle \,, \\
\lambda_2 &=& M \langle 00 \vert \hat{I}_c \otimes \hat{I}_h \vert 11 \rangle \,.
\ee
In the following, we will assume to be working at the symmetric point, \textit{i.e.} when the magnetic flux is exactly equal to half a flux quantum. At this point, it has been shown that the supercurrent operator $\hat{I}_k$ (with $k=c,h$) takes the simple form $b_k \hat{\sigma}_x^{(k)}$ with $b_k$ a real number \cite{Liu}. Here the Pauli matrix operator $\hat{\sigma}_x^{k}$ is defined as
\be
\hat{\sigma}_x^{k} = \vert 0 \rangle_k \langle 1 \vert + \vert 1 \rangle_k \langle 0 \vert \,.
\ee
When the two qubits are on resonance, $\omega_c = \omega_h \equiv E$, simple energy-scale arguments allows us to reduce the Hamiltonian (\ref{eq.Ham_flux}) to
\be
\label{eq.Ham_flux_red}
\hat{H}_{flux} &=& E ( \vert 1 \rangle \langle 1 \vert \otimes\mathds{1} + \mathds{1}\otimes \vert 1 \rangle \langle 1 \vert) + \lambda_1 (\vert 01 \rangle \langle 10 \vert + h.c. ) \,. \nonumber \\
&&
\ee
Equation (\ref{eq.Ham_flux_red})  is the exact analogue to Eqs.~\eqref{eq.absHo} and \eqref{eq.absHint} in the main text, where the interaction strength is given by $g \equiv \lambda_1 = M \bra{01}\hat{\sigma}_x^{(c)} \hat{\sigma}_x^{(h)} \ket{10}$.

Moreover, each qubit is coupled to a 'bath', represented by an external coil used for instance to generate the magnetic field enclosed by the superconducting loop. Each external circuit has a fluctuating current $i_k$ flowing through it, it is characterized by a finite impedance which is at the origin of dissipation processes. These dissipation processes are at the origin of the finite coherence time of the flux qubits for instance. %Indeed, the current $i_k$ flowing through this external coil is fluctuating. 
More precisely, the fluctuating current satisfies $\langle \hat{i}_k(t) \rangle =0$ and is characterized by a spectral density $S_k(E,T)$ which depends on the admittance $Y(E)$ \cite{Martinis03}
\be
\label{eq.noise}
S_k(E,T) = E \, \textrm{Re}[Y(E)] \left(  n_B (E,T) +1 \right) .
\ee 
Here $n_B$ denotes the Bose-Einstein distribution with a chemical potentials $E$ and temperature $T$,   $n_B(E,T) = 1/(e^{E/T}-1)$. The interaction Hamiltonian between the flux qubits and their own dissipative environment takes the form \cite{Martinis03, Chen}
\be
\hat{H}_{q-e}(t) = \sqrt{\Gamma_c}  \, \hat{i}_c(t) (\hat{\sigma}_x^{(c)} \otimes \mathds{1}) +  \sqrt{\Gamma_h} \,  \hat{i}_h(t) (\mathds{1} \otimes \hat{\sigma}_x^{(h)})\,.
\ee
By tuning the noise of the external circuit coupled to the flux qubit, one can therefore control the temperature of the environment.\\

The derivation of the master equation follows standard quantum optics calculations \cite{Breuer, Schaller}. The Hamiltonian of the total open quantum system reads
\be
\hat{H}_{tot} = \hat{H}_{flux} + \hat{H}_{q-e} + \hat{H}_E \,,
\ee
where $\hat{H}_E$ is the Hamiltonian of the environment which we do not need to specify here. Assuming a weak coupling between the qubits and their respective environment allows us to perform perturbation theory. The evolution of the total open quantum system is described by the von Neumann equation in the interaction picture (labelled by $(I)$)
\be
\dot{\rho}^{(I)}(t) = i \left[ \rho^{(I)}(t), \hat{H}_{q-e}^{(I)}(t) \right]\,,
\ee
with 
\be
\rho^{(I)}(t) &=& e^{i \hat{H}_{flux} t} \rho(t) e^{- i \hat{H}_{flux} t} \\
\hat{H}_{q-e}^{(I)}(t) &=& e^{i \hat{H}_{flux} t} \hat{H}_{q-e}(t)  e^{- i \hat{H}_{flux} t} \,.
\ee
Here $\rho(t)$ is the density operator of the total open quantum system. To derive a master equation in the Lindblad form, the dynamics of the open quantum system needs to satisfy several properties. First, one has to assume that the two external environments (external circuits with the magnetic coils) are large enough such that they remain unaffected by the presence of the qubits. This corresponds to the so-called Born approximation. In our model, the fluctuating currents through the external circuits do not depend on the qubits'  states which ensure this condition. Second, the bath correlation functions must decay rapidly compared to the dynamics of the qubits (the Markov assumption). The noise spectrum of each external circuit satisfies this condition. Neglecting fast oscillatory terms (secular approximation), we finally arrive at a master equation of the Lindblad type  
\begin{eqnarray}
\label{eq.fluxmaster}
\frac{\partial\rho}{\partial t}  = i [\rho,\hat{H}_{flux}] & + \sum_{i=1}^4 \Gamma_i^+  \left( \hat{J}_i \rho \hat{J}_i^\dagger - \frac{1}{2} \left\{ \hat{J}_i^\dagger \hat{J}_i, \rho \right\} \right) \nonumber \\
& +  \sum_{i=1}^4 \Gamma_i^-  \left( \hat{J}^\dagger_i \rho \hat{J}_i - \frac{1}{2} \left\{ \hat{J}_i \hat{J}_i^\dagger, \rho \right\} \right)\,.
\end{eqnarray}

The Lindblad operators $\hat{J}_i$ correspond to the four different processes by which the pair of qubits can receive one excitation from the baths. They are $\hat{J}_1 = \ketbra{0}{0}\otimes\hat{\sigma}_+$, $\hat{J}_2 = \hat{\sigma}_+\otimes\ketbra{0}{0}$, $\hat{J}_3 = \ketbra{1}{1}\otimes\hat{\sigma}_+$, $\hat{J}_4 = \hat{\sigma}_+\otimes\ketbra{1}{1}$. Their conjugates correspond to the inverse processes by which the qubit system looses one excitation. For instance, $\hat{J}_1$ corresponds to the hot qubit going from ground to excited state with the cold qubit in the ground state, and $\hat{J}_1^\dagger = \ketbra{0}{0}\otimes\hat{\sigma}_-$ corresponds to the hot qubit going from excited to ground state with the cold qubit in the ground state. The rates for these processes to occur are determined by the coupling probabilities, by the capacitance of the superconducting circuit and by the noise spectrum of each environment which is itself proportional to the Bose-Einstein distribution $n_B$\cite{Martinis03}.
\begin{eqnarray}
& \Gamma_1^+ =\Gamma_3^+ =   \Gamma_h \, n_B(E,T_h) ,  \nonumber \\
&  \Gamma_2^+ =\Gamma_4^+ = \Gamma_c \, n_B(E,T_c) , \\
& \Gamma_1^- = \Gamma_3^-  = \Gamma_h \, (1+ n_B(E,T_h)) , \nonumber \\ 
& \Gamma_2^- = \Gamma_4^- = \Gamma_c \, (1+ n_B(E,T_c)) , \nonumber
\end{eqnarray}
with E being set to 1. The coefficients $\Gamma_c$,$\Gamma_h$ take into account all parameters of the total circuit -- qubit and external coil. \figref{fig.plots_impl}(a) in the main text was obtained with $\Gamma_c$ ranging from $\sim 10^{-6}$ to $\sim 2\cdot 10^{-3}$, and $\Gamma_h$ in the range from $\sim 4\cdot 10^{-4}$ to $\sim 10^{-2}$. The rates satisfy the detailed balance equation $\Gamma^+_i / \Gamma^-_i = e^{-E/T_k}$.

\section{Implementation with a double quantum dot}
\label{app.impl2}

The second system we propose consists of two quantum dots weakly coupled through a tunnel barrier. This system is well known as a double quantum dot and has been widely investigated in the context of quantum transport experiments for its coherence properties \cite{Petersson10,Hanson07,Ihn09,delbecq,frey}. Although the double quantum dot traditionally plays the role of a single qubit, we consider here a different situation. Each dot can only by occupied by a single electron (we consider here spin-less electrons) and corresponds to a single qubit whose states correspond to the empty and occupied states, $\{ \ket{0}$, $\ket{1} \}$ . This system is highly tuneable with the help of external control gate voltages, which allows us  to operate this system as an efficient autonomous thermal machine. We consider for instance a large intra-dot Coulomb interaction to ensure single-occupancy of each dot and we assume a finite inter-dot Coulomb energy. When the two dots are on resonance (their eigenenergies are set to $E$), the Hamiltonian of this system is  similar to \eqref{eq.Ham_flux_red} with the energy $g$ setting the tunnel coupling between the two dots and an additional term characterised by the inter-dot Coulomb energy $U$
\begin{equation}
\label{eq.Ham_qd}
\hat{H}_{dot} = E \left(  \vert 1 \rangle \langle 1 \vert \otimes\mathds{1} + \mathds{1}\otimes \vert 1 \rangle \langle 1 \vert \right) + g (\vert 01 \rangle \langle 10 \vert + h.c. ) + U \ket{11}\bra{11} .
\end{equation}

Each dot $k = c,h$ is furthermore tunnel-coupled with an amplitude $\sqrt{\Gamma_k}$ to an independent fermionic reservoir characterised by a temperature $T_k$ and a chemical potential $\mu_k$. The energy $E$ of the two dots serves a reference for both reservoirs, \textit{i.e} $\mu_c = \mu_h \equiv E$. When the interaction between the dots and the reservoirs is weak, only single-charge tunnelling events occur with a probability proportional to $\Gamma_{c,h}$. The qubit-environement Hamiltonian takes the form
\begin{equation}
\hat{H}_{q-e} = \sqrt{\Gamma_c} \, \big( \hat{d}_c \left( \vert 1 \rangle_c \langle 0 \vert \otimes \mathds{1} \right) + h.c. \big) +  \sqrt{\Gamma_h} \, \big( \hat{d}_h \, \left( \mathds{1} \otimes \vert 1 \rangle_h \langle 0 \vert \right) + h.c. \big)\,.
\end{equation}
Here the operator $\hat{d}_k$ ($\hat{d}^\dagger_k$) is the fermionic annihilation (creation) operator for the reservoir $k$. The dynamics of this system is captured by a master equation identical to \eqref{eq.fluxmaster}, with the replacements $\hat{H}_{flux} \rightarrow \hat{H}_{dot}$ and $\Gamma^{\pm}_i \rightarrow \tilde{\Gamma}^{\pm}_i$. In contrast to the previous implementation where bosonic excitations were considered, the transition rates $\tilde{\Gamma}^{\pm}_i$ are now set by the Fermi-Dirac distribution $n_F$ to take into account the fermionic nature of the electrons transiting from one reservoir to the other through the two dots \cite{Schaller}
\begin{eqnarray}
&\tilde{\Gamma}_1^+ = \tilde{\Gamma}_3^+  = \Gamma_h \,  n_F(E,T_h) , \nonumber \\
&\tilde{\Gamma}_2^+ = \tilde{\Gamma}_4^+=  \Gamma_c \, n_F(E,T_c)  , \\
&\tilde{\Gamma}_1^- = \tilde{\Gamma}_3^- =  \Gamma_h \, (1-n_F(E,T_h)) , \nonumber \\
&\tilde{\Gamma}_2^+ = \tilde{\Gamma}_4^- = \Gamma_c \,  (1-n_F(E,T_c)). \nonumber
\end{eqnarray}
Here $n_F(E,T) = 1/(e^{E/T}+1)$ is the Fermi-Dirac distribution. In this case, \figref{fig.plots_impl}(b,c) of the main text corresponds to $\Gamma_c$ ranging from $\sim 6\cdot 10^{-5}$  to $\sim 10^{-2}$ and $\Gamma_h$ ranging from $\sim 7\cdot 10^{-5}$ to $\sim 10^{-2}$. Again, these rates are obtained following standard quantum optics calculations as presented for the flux-qubit case but can also be derived by applying the Fermi-golden rule to this system. We have also verified that they obey the detailed balance equation.%\\

\bibliographystyle{iopart-num}
%\bibliography{thermal_ent}

\providecommand{\newblock}{}

\end{document}